\documentclass[prc,aps,showpacs]{revtex4}
\usepackage{graphicx}
\begin{document}

\title{Production cross section of At radionuclides from  $^{7}$Li+$^{\textrm{nat}}$Pb and $^{9}$Be+$^{\textrm{nat}}$Tl reactions}
\author{Moumita Maiti\footnote{moumita.maiti@saha.ac.in}}
\affiliation{Chemical Sciences Division, Saha Institute of Nuclear Physics, 1/AF, Bidhannagar, Kolkata-700064, India.}
\author{Susanta Lahiri\footnote{susanta.lahiri@saha.ac.in (Reprint author), Fax: +91-33-2337-4637}}
\affiliation{Chemical Sciences Division, Saha Institute of Nuclear Physics, 1/AF, Bidhannagar, Kolkata-700064, India.}

\begin{abstract}

Earlier we reported theoretical studies on the probable production of astatine radionuclides from $^{6,7}$Li and $^{9}$Be-induced reactions on natural lead and thalliun targets, respectively. For the first time, in this report, production of astatine radionuclides has been investigated experimentally  with two heavy ion induced reactions: $^{9}$Be+$^{\textrm{nat}}$Tl and $^{7}$Li+$^{\textrm{nat}}$Pb. Formation cross sections of the evaporation residues, $^{207,208,209,210}$At, produced in (HI, xn) channel, have been measured by the stacked-foil technique followed by the off-line $\gamma$-spectrometry at the low incident energies ($<$50 MeV). Measured excitation functions have been explained in terms of compound nuclear reaction mechanism using Weisskopf-Ewing and Hauser-Feshbach model. Absolute cross section values are lower than the respective theoretical predictions.

\end{abstract}
\maketitle

\section{Introduction}

Astatine has now been more familier due to the potential application of $^{211}$At in targeted therapy. Owing to suitable nuclear propeties, $^{211}$At is promising in treating small tumor. Astatine radionuclides are produced artificially in the accelerator as the element has no naturally abundant isotope. Choice of suitable target-projectile combination and the knowledge of nuclear reaction data is therefore important in producing the radionuclide of choice. However, experimental cross section data is till date scare.

Ususlly, astatine radionuclides, $^{207-211}$At, are produced by bombarding $\alpha$-particle on natural bismuth target \cite{Hermanne,Henriksen,Schultz,Groppi}. Other production methods comprise $^{3}$He induced reaction on bismuth target \cite{Szucs,Vysotsky,Nagame} and high energy proton induced spallation reactions on heavy targets such as $^{238}$U, $^{232}$Th, etc. Sufficient amount of $^{207-211}$At radionuclides can also be produced by light heavy ion induced reactions, which are not well studied. A few reports dealt with the heavy ion induced production of astatine. Experimental measurement of formation cross sections of $^{208-211}$At produced through $^{7}$Li and $^{6}$Li induced reactions on enriched $^{208}$Pb target was reported in \cite{Wu,Hassan}. Production of $^{209,210}$At was reported in \cite{krsl, sl} where $^{7}$Li was bombarded on natural lead target aiming to study the chemical separation procedures of astatine from bulk lead.  The present authors also used $^{9}$Be-beam first time to produce  $^{208-210}$At from $^{\textrm{nat}}$Tl target and developed appropriate chemical separation method for production of t-radionuclides \cite{mmjrnc}. The encouraging yields of At-radionuclides in heavy ion activation prompted us to make theoretical investigation on the production possibility of astatine radionuclides through $^{\textrm{nat}}$Pb($^{7}$Li, xn), $^{\textrm{nat}}$Pb($^{6}$Li, xn), $^{\textrm{nat}}$Tl($^{9}$Be, xn) reactions \cite{mmslPRC}. The present report aims to measure the excitation functions of $^{207-210}$At produced in $^{7}$Li- and $^{9}$Be-induced reactions on natural lead and thallium targets, respectively and to validate our theoretical study \cite{mmslPRC}. 
Measured cross section data have been explained in terms of nuclear reaction mechanism comparating with two well established nuclear reaction model codes ALICE91 \cite{alice1,alice2} and PACE4 \cite{pace4}. Due to the limitation available accelerator facility, the report covers a small incident energy range. 

Section II describes the experimental procedure. Section III and IV deal with the data analysis and results of the present report.
 
\section{Experimental procedure}

Natural non-hygroscopic thallium carbonate, Tl$_{2}$CO$_{3}$ and lead nitrate, Pb(NO$_{3}$)$_{2}$, were used as target material. 
The targets of uniform thickness,  1.8$\pm$0.1 mg/cm$^{2}$ Tl$_{2}$CO$_{3}$ and 3.0$\pm$0.3 mg/cm$^{2}$ Pb(NO$_{3}$)$_{2}$, were prepared  by centrifugation technique on aluminium foil backing of thickness 1.5 mg/cm$^{2}$. Three such targets were mounted each time to prepare a target assembly,  which was then bombarded by the suitable projectile (e.g.$^{7}$Li or $^{9}$Be)  at the BARC-TIFR Pelletron Accelerator facility, Mumbai, India. The Tl$_{2}$CO$_{3}$ target stack was irradiated with a 47.6 MeV $^{9}$Be beam for 4.75 h up to a total charge of 388 $\mu$C and  Pb(NO$_{3}$)$_{2}$ target stack was irradiated with 46 MeV $^{7}$Li peojectile for 2.82 h up to a total charge of 1336 $\mu$C.  The residual products, if any, recoiled in the beam direction, were stopped in the aluminum backing. Large area of the catcher foils ensures the complete collection of recoiled evaporation residues. The beam intensity was measured in each experiment from the total charge collected in a electron suppressed Faraday cup stationed at the rear of the target assembly.

Irradiated foils were counted at the end of bombardment (EOB), to measure the $\gamma$-ray activity of the evaporation residues produced in the respective target matrix using an HPGe detector of 2.13 keV resolution at 1332 keV coupled with a PC based MCA. Each foil was counted in a regular time interval untill the complete decay of the residues. Use of centrifuged targets on aluminum backing restricted to measure separately the recoiled activity induced in the aluminum foils. However, in the present case, recoiled activity in the backing, if any, is expected to be negligible as it deals with low projectile energy. 

Beam energy degradation in the target and the catcher foils was calculated using the Stopping and Range of Ions in Matter (SRIM) \cite{srim}. Projectile energy at the target is the average of incident and outgoing beam energy. Energy loss is about  2\% in the thallium carbonate and lead nitrate targets. Product yields of the evaporation residues in each foil were calculated from the background subtracted peak area count correspond to a particular $\gamma$-ray energy. The nuclear spectroscopic data of the radionuclides studied in this article is enlisted in the Table \ref{mmt1} \cite{nudat2}. The cross sections of the evaporation residues produced at various incident energies were calculated from the standard activation equation. The total associated error related to the cross section measurement was determined considering all the probable uncertainties and the data presented up to 95\% confidence level. A detail description of the calculation is available elsewhere \cite{mmTc}.

\section{Analysis of cross section}

In order to compare the measured cross sections, theoretical cross sections of $^{207-210}$At were calculated from $^{7}$Li+$^{\textrm{nat}}$Pb and $^{9}$Be+$^{\textrm{nat}}$Tl reactions using the nuclear reaction model codes PACE4 \cite{pace4} and ALICE91 \cite{alice1, alice2}. 

The code \textsc{PACE4} \cite{pace4} is the modified version of  \textsc{PACE} (Projection Angular momentum Coupled Evaporation) \cite{pace} working in the framework of \textsc{LISE}++ \cite{nscl} with several new features. It uses Hauser-Feshbach model to follow the deexcitation of the excited nuclei. The transmission coefficients for light particle emission have been determined from the optical model potential with default optical model parameters. The code internally decides level densities and masses it needs during deexcitation. The Gilbert-Cameron level density prescription is used in the present work with $\textit{a}$, level density parameter, equals to A/9 MeV$^{-1}$. The ratio of $a_{f}$/$a_{n}$ is chosen as unity. Fission is considered as a decay mode where finite range fission barrier of Sierk has been used.  The compound nuclear fusion cross section is determined by using the Bass method. The yrast parameter is taken as unity.

The excitation functions of $^{207-210}$At, have been calculated using the code ALICE91 \cite{alice1,alice2} with geometry dependent hybrid model \cite{alice2} for preequilibrium emissions and Weisskopf-Ewing formalism for equilibrium emissions. A separate calculation has also been done using ALICE91 only with Weisskopf-Ewing model option for the excitation functions of  $^{208-211}$At. The details of hybrid model is available in our previous papers \cite{mmslPRC,mmprc}.
The emission of light particles, $n$, $p$, $d$ and $\alpha$, are considered from the residual nuclides of 12 mass unit wide and 10 charge unit deep including the composite nucleus. Fermi gas level density has been used for the calculation of reaction cross sections. Reverse channel reaction cross sections have been calculated using the optical model. The level density parameter, $a$ is taken as A/9, the default value for the code. Rotating finite range fission barriers of Sierk has been chosen. Total number of nucleons in the projectile has been chosen as the initial exciton number for the preequilibrium emission calculation.

Formation cross sections of the residues were calculated separately from  $^{7}$Li- and $^{9}$Be-induced reactions on each naturally occurring isotope of Pb and Tl, respectively, and the total formation cross section was calculated taking the weighted average of all the naturally occurring isotopes.

\section{Results and discussion}

\begin{figure}
\begin{center}
\includegraphics[height=8.0cm]{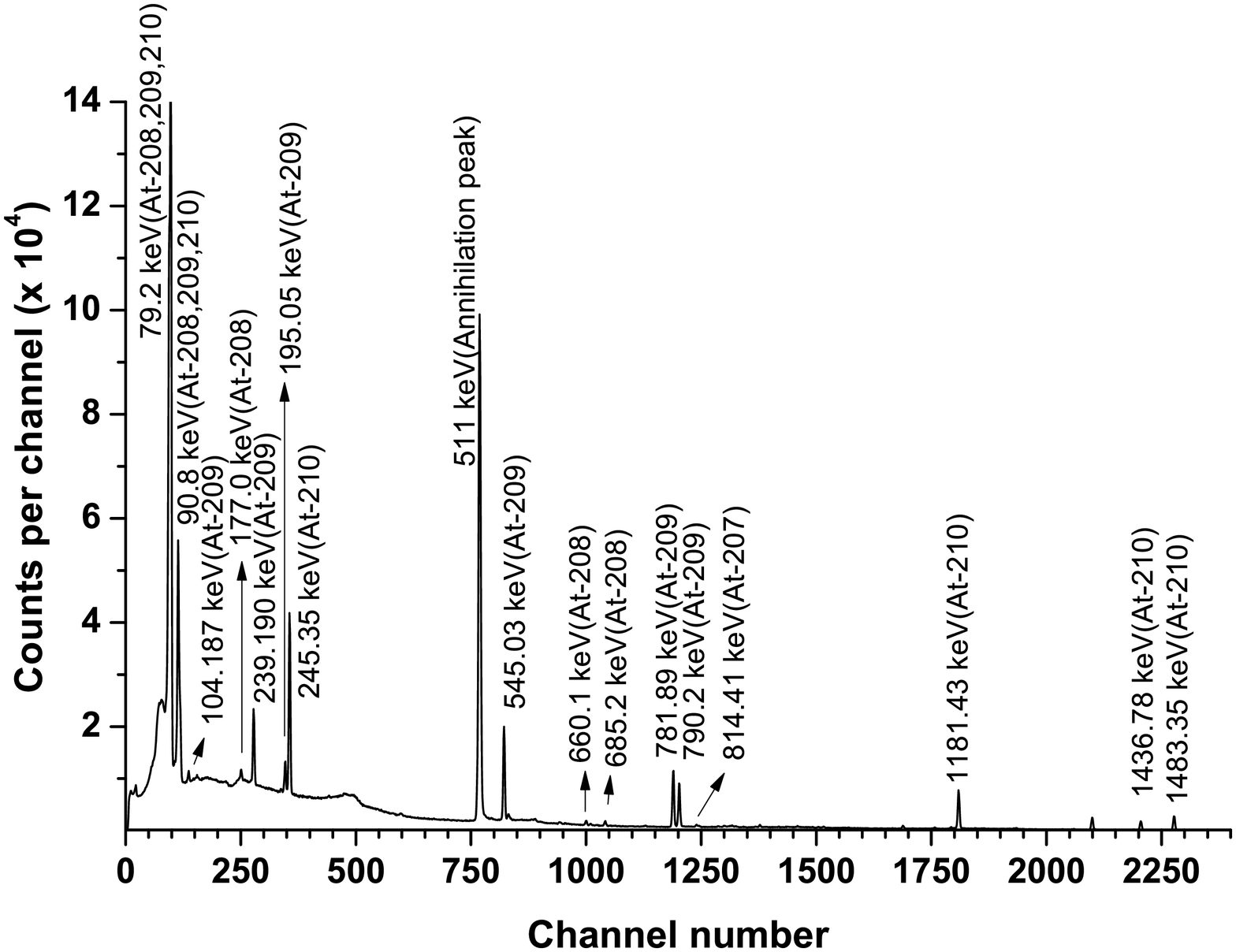}
\caption{$\gamma$-ray spectrum of the radionuclides produced in $^{7}$Li+$^{\textrm{nat}}$Pb reaction at 46 MeV incident energy after 1.5 h of EOB.} 
\label{F1}
\end{center}
\end{figure}

\begin{figure}
\begin{center}
\includegraphics[height=8.0cm]{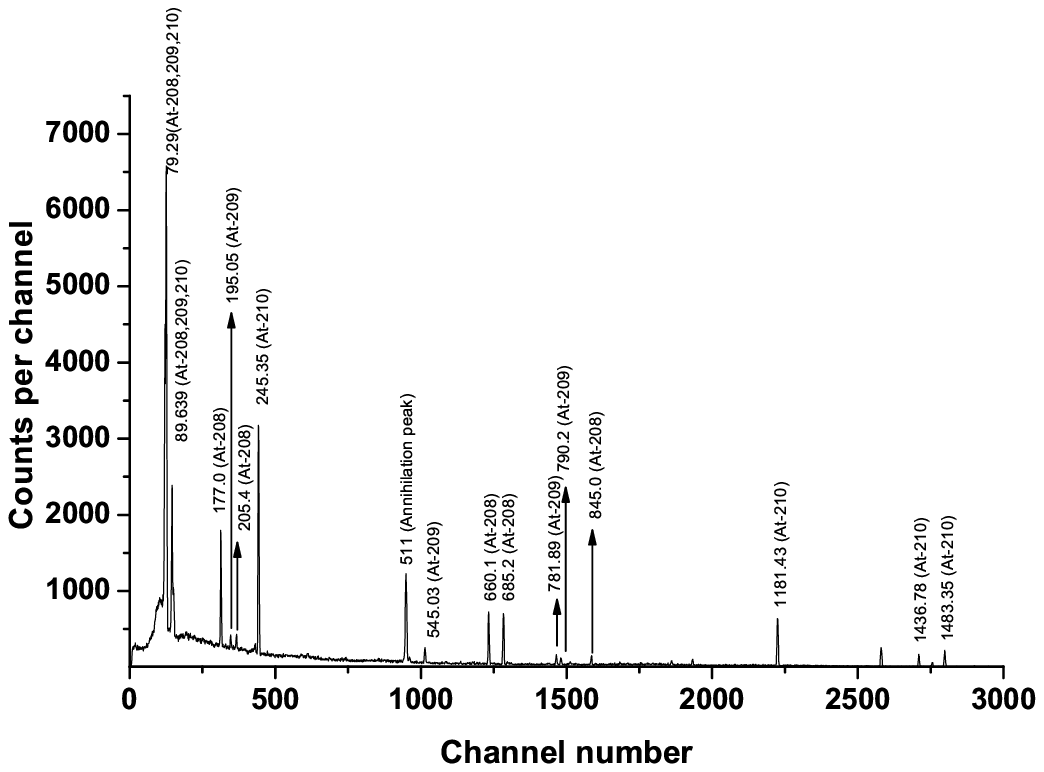}
\caption{$\gamma$-ray spectrum of the radionuclides produced in $^{9}$Be+$^{\textrm{nat}}$Tl reaction at 47.6 MeV incident energy after 2 h of EOB.} 
\label{F2}
\end{center}
\end{figure}

Analysis of $\gamma$-spectra collected at different time intervals after EOB assured the production of various proton rich astatine radionuclides, $^{207, 208, 209, 210}$At, in the target matrix due to the bombardment of $^{7}$Li- and $^{9}$Be-projectiles on $^{\textrm{nat}}$Pb and $^{\textrm{nat}}$Tl targets, respectively, at low incident energies ($<$50 MeV). Figure \ref{F1} and \ref{F2} represent $\gamma$-spectrum of the $^{7}$Li- and $^{9}$Be-irradiated $^{\textrm{nat}}$Pb and $^{\textrm{nat}}$Tl targets at the highest incident energies, 46 MeV and 47.6 MeV, respectively. The astatine radionuclides produced in the particular target-projectile combination is tabulated in Table \ref{mmt1} along with the reaction threshold values.
Theoretical investigation \cite{mmslPRC} shows considerable possibility of producing  $^{211}$At ($\approx$ 400 mb) in $^{7}$Li induced reaction on $^{\textrm{nat}}$Pb. However, it was not possible to identify $^{211}$At by $\gamma$-ray spectrometry in the present experimentdue due to its low intensity $\gamma$-ray emissions.

\begin{figure}
\begin{center}
\includegraphics[height=8.0cm]{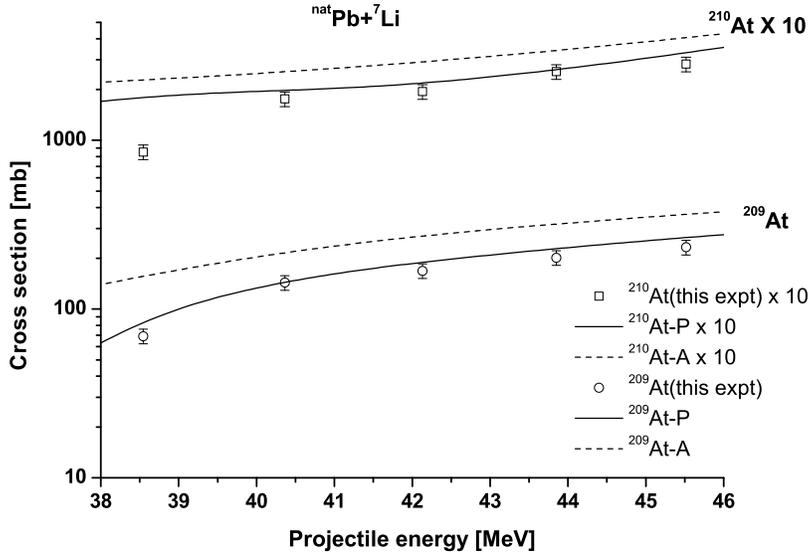}
\caption{Comparison between measured cross sections of $^{210}$At and $^{209}$At from $^{7}$Li+$^{\textrm{nat}}$Pb reactions and that theoretically predicted from \textsc{PACE4} and \textsc{ALICE91}. -P stands for \textsc{PACE4} and -A stands for \textsc{ALICE91}.}
\label{fig1}
\end{center}
\end{figure}

\begin{figure}
\begin{center}
\includegraphics[height=8.0cm]{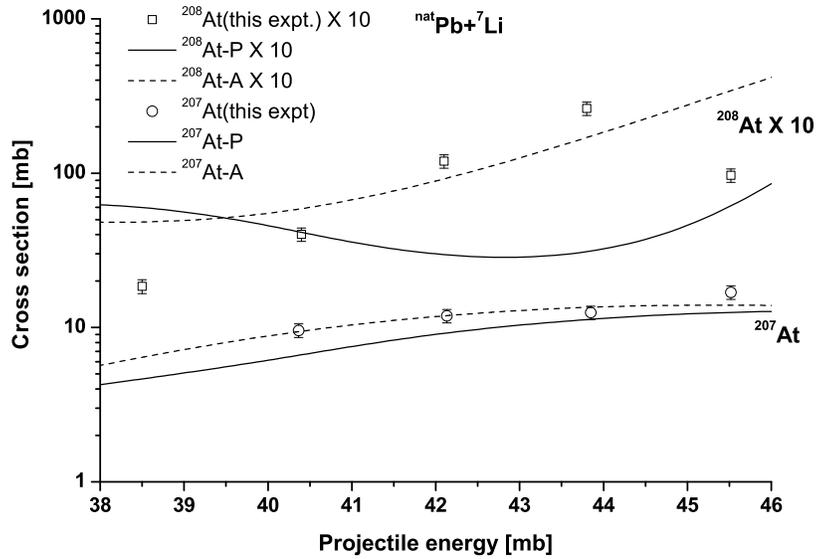}
\caption{Same as Fig. \ref{fig1} for $^{208}$At and $^{207}$At} 
\label{fig2}
\end{center}
\end{figure}

Cross sections measured for $^{207-210}$At from $^{7}$Li+$^{\textrm{nat}}$Pb vreaction in 46-38 MeV projectile energy and $^{208-210}$At  from $^{9}$Be+$^{\textrm{nat}}$Tl reaction in 47.2-42 MeV have been compared with theoretical predictions of \textsc{PACE4} and \textsc{ALICE91} as shown in Figs. \ref{fig1}-\ref{fig2}. Though ALICE91 takes care of the preequilibrium emissions, it has been observed in general that preequilibrium reaction has no contribution, even in the highest projectile energy in both the cases. As a result, the comparison is practically between the two compound nuclear reaction models Weisskopf-Ewing and Hauser-Feshbach. 

\begin{figure}
\begin{center}
\includegraphics[height=8.0cm]{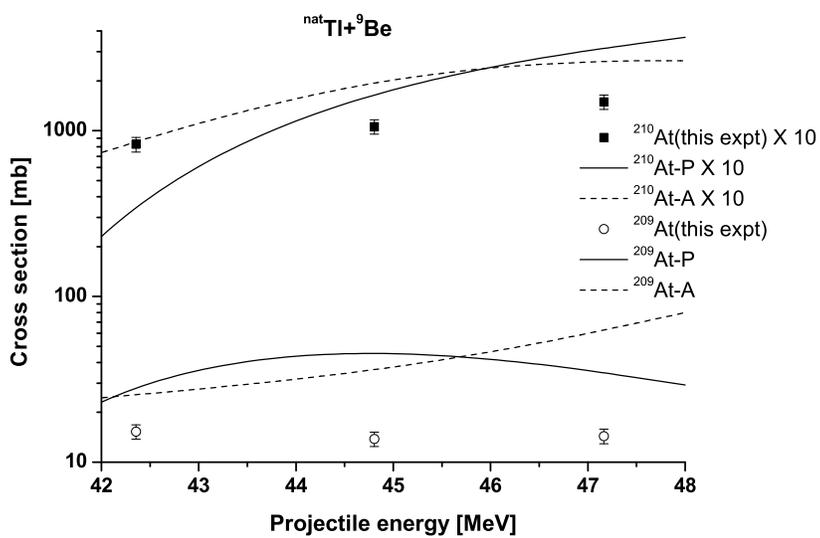}
\caption{Same as Fig. \ref{fig1} for $^{210}$At and $^{209}$At from $^{9}$Be+$^{\textrm{nat}}$Tl reaction} 
\label{fig3}
\end{center}
\end{figure}

\begin{figure}
\begin{center}
\includegraphics[height=8.0cm]{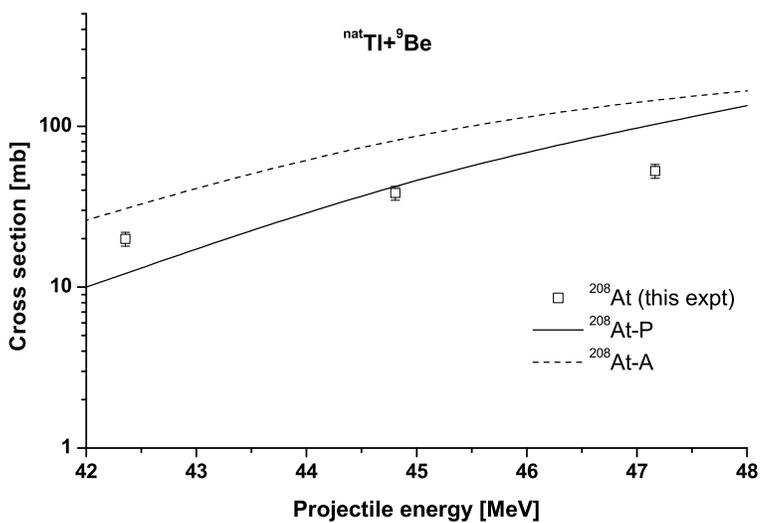}
\caption{Same as Fig. \ref{fig3} for $^{208}$At } 
\label{fig4}
\end{center}
\end{figure}

It is observed from Fig. \ref{fig1} that experimental excitations of $^{210}$At $^{209}$At are well evaluated by PACE4 whereas ALICE91 overpredicts the data $\approx$ 40\%. Both the theoretical predictions agree with the measured cross sections for $^{207}$At, but they neither reproduce measured cross section nor the trend for $^{208}$At (Fig. \ref{fig2}). Similar phenomenon was observed in case of $^{9}$Be+$^{\textrm{nat}}$Tl reaction (Figs. \ref{fig3} - \ref{fig4}). PACE4 calculation underpredicts the measured data at the lowest incident energy while it overpredicts the measured data at higher energies for $^{210}$At and $^{208}$At. ALICE91 also overpredicts the measured data at higher energies but agrees well with the cross section values at 42.3 MeV for $^{210}$At and $^{208}$At, respectively. However, no agreement was found between theory and experiment for $^{209}$At and the measured cross sections $^{209}$At are almost constant in 42-47.5 MeV projectile energy range. It was critical to make any definite comment on the cross section data as a small incident energy range of the expected excitation functions shown in \cite{mmslPRC} was covered in the present report. However, analysis of the measured data reveals the signature of compound nuclear reaction in producing $^{207-210}$At  in the reported incident energy region. It has been experienced experimentally that production of $^{208}$At is higher than that of $^{209}$At in this energy range. The fact is in well agreement with theoretical evaluation. It is known that Weisskopf-Ewing model gives higher cross section values than the Hauser-Feshbach model as it sacrifices rigor of physics. This fact is also reflected in the comparison shown in the figures.

\section{Conclusion}

This work reports first time the measured production cross sections of $^{207,208,209,210}$At from two separate heavy ion induced reactions, $^{7}$Li+$^{\textrm{nat}}$Pb and $^{9}$Be+$^{\textrm{nat}}$Tl, respectively. 
Production cross sections of the astatine radionuclides, which are expected to be evaporation residues, have been compared with two established evaporation models: Weisskopf-Ewing and Hauser-Feshbach model. Measured cross sections are in general lower than the theoretical expectations. The present work is limited due to the available projectile energy and reports only a small part of the total excitation functions of the evaporation residues. However, the measured cross section data are important to validate the theoretical predictions reported in \cite{mmslPRC} and to enrich the nuclear reaction data bank towards the production of various proton rich astatine radionuclides.

\begin{acknowledgments}

Authors are thankful to target laboratory VECC, Kolkata, for preparing targets. Thanks to pelletron staff of BARC-TIFR pelletron facility, Mumbai, for their cooperation and help during experiment. M. Maiti expresses sincere thanks to the Council of Scientific and Industrial Research (CSIR) for providing necessary grants. This work has been carried out as part of the SINP-DAE, XI five year plan project "Trace Analysis: Detection, Dynamics and Speciation (TADDS)".  

\end{acknowledgments}

\begin{table}[h]
\caption{Nuclear spectrometric data \cite{nudat2} of the radionuclides produced through different nuclear reactions.}
\label{mmt1}
\begin{tabular}{cccccccc}
\hline
   Isotope & T$_{1/2}$ & Decay mode(\%) & E$_{\gamma}$keV(I$_{\gamma}$\%) & $^{7}$Li+$^{\textrm{nat}}$Pb & E$_{th}$(MeV) & $^{9}$Be+$^{\textrm{nat}}$Tl &  E$_{th}$(MeV) \\
\hline 
 $^{210}$At & 8.1 h & $\epsilon$(99.82)$\alpha$(0.18) & 1181.43(99) & $^{208}$Pb($^{7}$Li,$5n$)& 36.41 & $^{203}$Tl($^{9}$Be, $2n$) & 19.41 \\
 						&				&																	&							& $^{207}$Pb($^{7}$Li,$4n$)& 28.80 & $^{205}$Tl($^{9}$Be, $4n$) & 34.23 \\
 						&				&																	&							&	$^{206}$Pb($^{7}$Li,$3n$)& 21.84 &														&				\\
 						
$^{209}$At & 5.41 h & $\epsilon$(95.9)$\alpha$(4.1)  & 545.03(91) & $^{207}$Pb($^{7}$Li,$5n$)& 36.21 & $^{203}$Tl($^{9}$Be, $3n$) & 26.89 \\
 					 &				&																 &						& $^{206}$Pb($^{7}$Li,$4n$)& 29.25 & $^{205}$Tl($^{9}$Be, $5n$) & 41.71 \\
 					 &				&																 &						&	$^{204}$Pb($^{7}$Li,$3n$)& 13.93 &														&				\\
 					 
$^{208}$At & 1.63 h & $\epsilon$(99.45)$\alpha$(0.55) & 686.527(97.6) & $^{206}$Pb($^{7}$Li,$5n$)& 37.99 & $^{203}$Tl($^{9}$Be, $4n$) & 35.73 \\					
 					 &				&																 &						&	$^{204}$Pb($^{7}$Li,$3n$)& 22.68 &														&				\\
$^{207}$At & 1.8 h & $\epsilon$(91.4)$\alpha$(8.6) & 814.41(45) & $^{204}$Pb($^{7}$Li,$3n$)& 30.25		&														&  			\\
 
\hline
\end{tabular}
\end{table}

\end{document}